\documentclass[doublecol]{epl2} 
\usepackage{graphicx}
\usepackage{epstopdf}
\usepackage{cite}
\usepackage{amssymb,amsfonts,amsmath}

\title{Undulatory swimming in fluids with polymer networks}

\author{D.A. Gagnon\inst{1} \and X.N. Shen\inst{1} \and P. E. Arratia\inst{1}}
\shortauthor{D.A. Gagnon \etal}

\institute{                    
  \inst{1} University of Pennsylvania - Department of Mechanical Engineering and Applied Mechanics, Philadelphia PA 19104\\
}

\abstract{
The motility behavior of the nematode \textit{Caenorhabditis elegans} in polymeric solutions of varying concentrations is systematically investigated in experiments using tracking and velocimetry methods. As the polymer concentration is increased, the solution undergoes a transition from the semi-dilute to the concentrated regime, where these rod-like polymers entangle, align, and form networks. Remarkably, we find an enhancement in the nematode's swimming speed of approximately 65\% in concentrated solutions compared to semi-dilute solutions. Using velocimetry methods, we show that the undulatory swimming motion of the nematode induces an anisotropic mechanical response in the fluid. This anisotropy, which arises from the fluid micro-structure, is responsible for the observed increase in swimming speed.}

\begin{document}

\maketitle

\section{Introduction} Much attention has been devoted to the problem of swimming and self-propulsion in Newtonian fluids such as water and oils~\cite{Brennen1977, Rodenborn2013, Drescher2011, Guasto2010, Lauga2009, Gray1955}. However, many fluids of practical interest contain solids and/or polymers which form structures at an intermediate scale between the size of a molecule and the macroscopic length scale of the system~\cite{Larson1999}. Such ``complex fluids'' often display non-Newtonian rheological behavior including shear-thinning viscosity and viscoelasticity. The interplay between the fluid's internal structure (e.g. polymer networks) and self-propulsion is critical to many biological processes such as reproduction~\cite{Guzick2001}, bacterial infection~\cite{Josenhans2002}, and bio-degradation in soil~\cite{Alexander1991}. For example, the bacteria \emph{H. pylori} can modulate the acidity of gastric mucus and thus untangle glycol-protein networks, reducing the resistance of mucus~\cite{Celli2009}, and nematodes can burrow through the networks present in wet soil aiding in soil aeration, water storage, and soil fertility~\cite{Juarez2010, Jung2010}. Understanding swimming in complex fluidic environments is thus pertinent to the treatment of human diseases as well as the characterization and maintenance of ecological systems.

Experimental observations have revealed that polymer networks can enhance the swimming speed of flagellated bacteria moving in solutions containing long-chain polymer molecules ~\cite{Berg1979, Schneider 1974}. For these small organisms ($L<10$~$\mu$m), it has been argued that the main mechanism for this propulsion enhancement is due to the benefits of pushing against a quasi-rigid polymer network~\cite{Berg1979,Magariyama2002}. The role of the mechanical properties of fluid internal networks on an organism's swimming behavior has also been investigated in numerical~\cite{Fu2010, Du2012,Magariyama2002, Nakamura2006} and theoretical \cite{Leshansky2009} studies.  Numerical studies of swimming in fluids structured with polymer networks resembling natural environments have postulated that the shapes and dynamics of internal networks are accounted for by two effective anisotropic viscosities~\cite{Magariyama2002, Nakamura2006}, which qualitatively explain some of the observed propulsion enhancement~\cite{Berg1979, Schneider1974}. Such anisotropic viscosities, however, are difficult to measure and apply to quantitative analysis. In heterogeneous, gel-like environments, modeled by embedding stationary objects in an incompressible viscous fluid, the swimming speed of a microorganism can be enhanced by the underlying structures in the fluid~\cite{Leshansky2009}. For internal networks made of small molecules, such as a binary blend of two intermixed fluids, a two-fluid model predicts an enhancement in swimming speed for stiff and compressible networks~\cite{Fu2010}, and a reduction in swimming speed when local distributions of volume fractions of the two phases scale differently for thrust and drag~\cite{Du2012}. Overall, the observed propulsion speed variations in these studies underscore the important role of the fluid structures on the swimming behavior of microorganisms.
\begin{figure}[t] 
    \begin{center}
      \includegraphics[width=.47\textwidth]{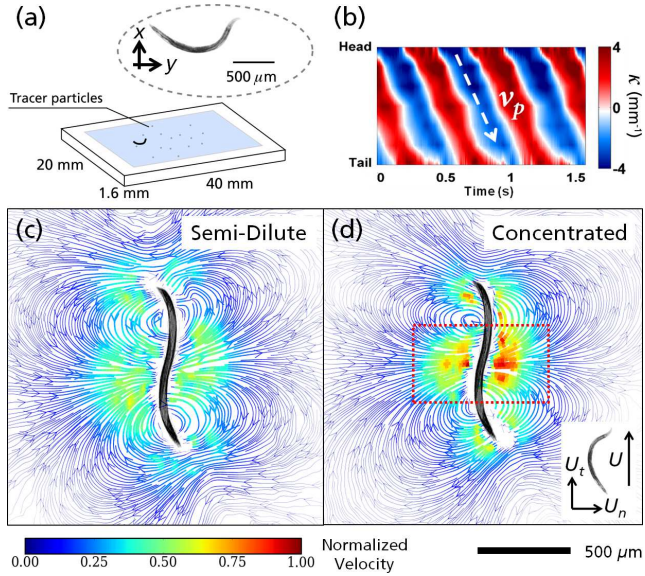}
           \caption{(Color Online) {\bf{(a)}} Schematic of nematode \emph{C. elegans} swimming in a sealed fluidic chamber. {\bf{(b)}} Nematode's centerline and body curvature during swimming. Curvature plot for approximately 3 beating cycles reveals traveling wave propagating from head to tail.  Wave speed illustrated by the white dashed line with arrow indicating the direction. {\bf{(c, d)}} Flow streamlines color-coded by normalized velocity magnitude during swimming for 2000 ppm (semi-dilute) and 4000 ppm (concentrated) cases. The red box highlights a region of high relative flow in the concentrated regimes; inset defines normal and tangential components relative to the worm body. \label{figure1}}
    \end{center}
\end{figure}

In this manuscript, we investigate the effects of polymer networks on the swimming dynamics of the nematode \emph{Caenorhabditis elegans} in experiments using tracking and velocimetry methods. Polymer networks are formed by controlling the concentration of the bio-compatible rod-like polymer xanthan gum in water.  We find an enhancement of approximately 65\% in the nematode's swimming speed in concentrated polymeric solutions compared to semi-dilute solutions. Due to the relatively large size of the nematode ($L \approx 1$~mm) compared to the polymer networks ($\sim 10~\mu$m), the mechanism of pushing against a quasi-static polymer network is insufficient to explain the increase in swimming speed. We argue that the propulsion enhancement arises from local shear-induced anisotropy.\\
\section{Experimental Methods} 
 Experiments are performed in an acrylic fluidic chamber (Fig.~\ref{figure1}\emph{a}) that is 20~mm wide, 40~mm long and 1.6~mm deep. All experiments are performed with wild-type, adult \emph{C. elegans} that are on the average 1 mm in length and 80~$\mu$m in diameter. The nematodes swimming motion is imaged using a microscope with a depth of field of 30~$\mu$m and a fast camera operated at 100 frames per seconds in order to accurately resolve small body displacements. To minimize boundary effects from the top and bottom walls, the imaging plane is focused at the center of the chamber. All data presented here pertain to nematodes swimming at the center plane of the fluidic channel. Out-of-plane recordings are discarded. An average of 15 nematodes are recorded for each experiment (see movie in SM).

Figure~\ref{figure1}\emph{a} shows a sample snapshot of a nematode swimming in the water-like buffer solution M9 (see sample movie in SM). Here, swimming speed $U$ is calculated by differentiating the nematode's centroid position over time; we find $U \approx$~0.35~mm/s for nematodes swimming in M9 solutions.  The nematode's Reynolds number $Re=\rho UL/\mu \approx 0.2$, where $L$ is the nematode's length (1~mm), $\rho$ is fluid density, and $\mu$ is the fluid viscosity ($1$~mPa$\cdot$s). The nematode's shape-line in each frame is extracted by in-house software \cite{Krajacic2012} to calculate the nematode's body curvature, defined as $\kappa =d\phi/ds$. Here, $\phi$ is the angle made by the tangent to the $x$-axis in the laboratory frame at each point along the body centerline, and $s$ is the arc length coordinate from the head of the nematode ($s=0$) to its tail ($s=L$). Fig.~\ref{figure1}\emph{b} shows the spatio-temporal evolution of the nematode's body curvature $\kappa(s, t)$ for approximately three swimming cycles. This curvature plot shows the existence of periodic, well-defined, and diagonally oriented lines characteristic of bending waves propagating in time along the nematode body length, from which important kinematic quantities such as beating frequency $f$ and wave-speed $\nu_p$ are obtained~\cite{Sznitman2010Viscosity}; wave-speed is illustrated using a white dashed line with an arrow indicating the direction of waves in Fig.~\ref{figure1}\emph{b}. We find that $f \approx 2$~Hz and $\nu_p=5$~mm/s for nematodes swimming in M9 buffer solution. 

\begin{figure}[t] 
    \begin{center}
      \includegraphics[width=0.47\textwidth]{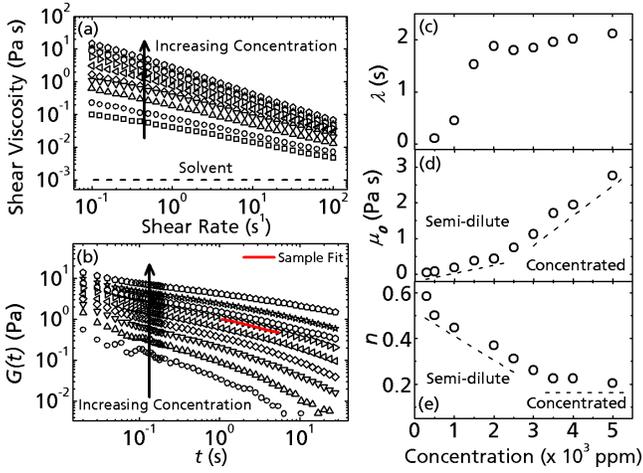}
     \caption{{\bf{(a)}} Shear viscosity of xanthan gum solutions showing power law behavior. {\bf{(b)}} Stress relaxation data for XG solutions fit to a linear viscoelastic model. {\bf{(c)}} Relaxation time $\lambda$, {\bf{(d)}} viscosity factor $\mu_{0}$, and {\bf{(e)}} power law index $n$ as a function of concentration. Two different trends for $\mu_{0}$ and $n$ indicate the structural transition of solutions from the semi-dilute to concentrated regime.\label{figure2}}
    \end{center}
\end{figure}

Polymeric fluids are prepared by adding small amounts of xanthan gum (XG) to deionized water. Xanthan gum is a semi-rigid rod-like polymer with a molecular weight of $2\times10^{6}$ Da (Sigma Aldrich, G1253). Polymer concentration ranges from 300 ppm to 5000 ppm by weight. The rheological properties of all polymeric solutions are characterized by a stress-controlled rheometer (RFS3, TA Instruments). Fig.~\ref{figure2}\emph{a} shows the shear viscosity $\mu$ of all fluids as a function of shear rate $\dot{\gamma}$. All fluids exhibit shear-thinning viscosity with the exception of M9 buffer solution which exhibits water-like behavior. This non-Newtonian viscosity behavior is well captured by an empirical power law fluid model of the type $\mu=\mu_{0}|\dot{\gamma}|^{n-1}$, where $\mu_{0}$ is the viscosity factor and $n$ is the power law index. 
Fluid relaxation times are obtained by fitting the stress relaxation data with the generalized linear viscoelastic model of a single relaxation time of the type $G(t) = G_0e^{-t/\lambda}$, where $G(t)$ is the fluid shear modulus and $\lambda$ is the longest relaxation time of the fluid. Figure~\ref{figure2}\emph{b} shows the data and a sample fitting and the values of $\lambda$ are shown in Fig.~\ref{figure2}\emph{c}. Figure~\ref{figure2}\emph{d} shows the values of $\mu_{0}$ as a function of XG polymer concentration. The different slopes found in Fig.~\ref{figure2}\emph{d} represent two distinct polymer concentration regimes, namely the semi-dilute and concentrated regimes. The transition from the semi-dilute to the concentrated regime seems to happen near or around 3000 ppm and it is commonly interpreted as a structural transition~\cite{Rodd2000, Doi1988}.  In concentrated solution, the shape and dynamic properties of polymer networks dominate flow behaviors; in semi-dilute solution, the hydrodynamic interactions among individual polymers dominate flow behaviors~\cite{Doi1988}. We note that the power law index $n$ of xanthan gum solutions also shows two regimes (Fig.~\ref{figure2}\emph{e}) due to the aforementioned structural transition~\cite{Wyatt2009}. We note that there is no expectation for concentration transition to be reflected by the change in slope of relaxation time $\lambda$ in Fig.~\ref{figure2}\emph{c}.

Particle tracking methods are used to measure the velocity fields generated by the nematodes swimming in the various polymeric fluids. (Sample movie available in SM). Figure~\ref{figure1}\emph{c}~and~\emph{d} show normalized velocity fields for \emph{C. elegans} swimming in semi-dilute (2000 ppm) and concentrated (4000 ppm) XG solutions. Velocities for each solution are normalized by their respective root-mean-square velocities and then scaled by the maximum between the two regimes. These normalized streamlines are color-coded to highlight regions of high (red) and low (blue) flow. We find large recirculating regions around the nematode's body and regions with large extensional (or straining) components. The flow structures are relatively similar for both cases, but we find differences around the nematode's midsection, where relative velocities in the concentrated case are nearly twice as large as those in the semi-dilute case and are predominately in the normal direction. We will discuss this feature later in the manuscript.

\section{Polymer Concentration Effects} 
\begin{figure}[h] 
    \begin{center}
      \includegraphics[width=0.47\textwidth]{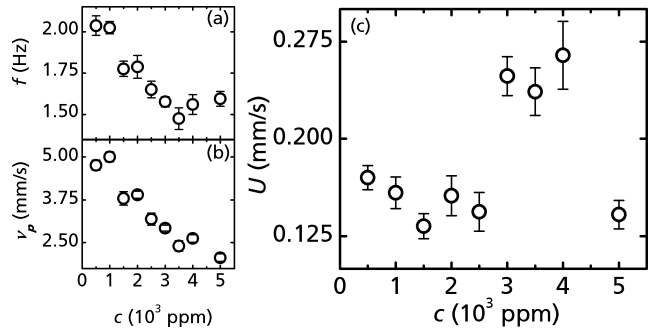}
      \caption{{\bf{(a, b)}} Swimming frequency $f$ and wave speed $\nu_p$ as a function of concentration. {\bf{(c)}} Swimming speed as a function of concentration. Swimming speed exhibits a rapid increase as the solution enters the concentrated regime.\label{figure3}}
    \end{center}
\end{figure}

We begin by analyzing the swimming kinematics of \emph{C. elegans} in XG solutions as a function of polymer concentration. Figures~\ref{figure3}\emph{a}-\emph{c} show the nematode's  beating frequency $f$, wave speed $\nu_p$, and swimming speed $U$ as a function of concentration, respectively.  As the polymer concentration increases, we find a gradual and monotonic decay in the nematode's wave speed $\nu_p$ and a decay in beating frequency $f$ below 3000 ppm followed by a plateau at higher concentrations (Fig.~\ref{figure3}\emph{a}~and~\emph{b} respectively). This decrease in such kinematic functions is expected as a result of increased viscosity as polymer concentration increases (Fig.~\ref{figure2}\emph{d}). 

Figure~\ref{figure3}\emph{c} shows that the nematode's swimming speed $U$ remains relatively constant for polymer concentrations below 3000 ppm. Surprisingly, however, we find a sudden increase in $U$ for concentrations above 3000 ppm. The values of $U$ are maintained around 0.15 mm/s in semi-dilute solutions but they quickly rise by 65\% to about 0.25 mm/s in concentrated solutions despite a significant increase in solution viscosity.  As expected, the swimming speed ultimately decreases as the concentration is further increased due to the nematode's finite power output~\cite{Shen2011}.  An increase in $U$ with viscosity has been previously reported for microorganisms moving in structured gel-like media, but the mechanisms are still not well understood~\cite{Berg1979, Schneider1974}. A recent theoretical work suggests that such increase may be due to the presence of polymer networks in the media and that microorganisms may be able to push against such quasi-static networks and move more efficiently~\cite{Magariyama2002}. However, because of the large difference in length scales between the nematode ($\approx 1$~mm) and the polymer networks ($\approx 10$~$\mu$m) as well as the lack of quasi-static flow fields in the concentrated regime (Fig.~\ref{figure1}\emph{d}), this notion does not adequately explain the observed propulsion enhancement.  In what follows, we will show that the propulsion enhancement for \emph{C. elegans} swimming in concentrated polymer solutions is related to shear-induced fluid anisotropy. 
\section{Swimming Speed \& Fluid Rheology} 
\begin{figure}[t] 
    \begin{center}
      \includegraphics[width=0.47\textwidth]{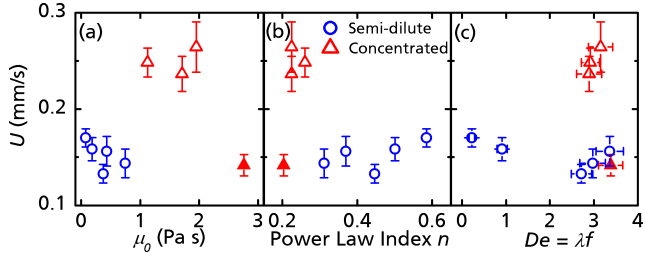}
       \caption{(Color Online) Nematode's swimming speed as a function of {\bf{(a)}} viscosity factor, {\bf{(b)}} power law index, {\bf{(c)}} and Deborah number or fluid elasticity (see text). \label{figure4}}
    \end{center}
\end{figure}

To better understand the mechanisms governing of this unusual enhancement in $U$, we investigate the effects of both fluid viscosity and elasticity on the nematode's swimming speed. Figure~\ref{figure4}\emph{a} and \emph{b} show the values of $U$ as a function of fluid viscosity coefficient $\mu_{0}$ and power law index $n$. Note that the transition from semi-dilute to concentrated regime in the XG solutions occurs at $\mu_{0} \approx 1$ Pa$\cdot$s and $n \approx 0.3$, as shown in Fig.~\ref{figure2}\emph{d} and \emph{e}.  Results show that the nematode is able to maintain a constant swimming speed ($\approx$ 0.15 mm/s) for $\mu_0 \lesssim 1$ Pa$\cdot$s but exhibits a rapid increase to $U \approx$ 0.25 mm/s for $\mu_0 \gtrsim 1$ Pa$\cdot$s (Fig.~\ref{figure4}\emph{a}). This corresponds to an increase in swimming speed of about 65$\%$ as the fluid viscosity increases. As expected, further increase in viscosity beyond~$\approx 2$~Pa$\cdot$s reduces swimming speed, which suggests that the enhancement in swimming speed is bounded by the nematode's limited power output~\cite{Shen2011}. The effects of shear thinning on the nematode's swimming speed (Fig.~\ref{figure4}\emph{b}) echo the similar pattern found with $\mu_{0}$. That is, the swimming speed is abruptly enhanced for XG solutions corresponding to $n \gtrsim 0.3$; below such value, the nematode's swimming speed remains relatively constant. 

Due to the relatively high polymer concentration and the formation of polymer networks, viscoelastic effects are expected in semi-dilute and concentrated XG solutions. Fluid elasticity is known to strongly affect the swimming behavior of microorganisms~\cite{Fu2009, Lauga2009,Shen2011}. For the case of \emph{C. elegans}, it has been recently shown the swimming speed decreases as fluid elasticity increases \emph{in dilute solutions}~\cite{Shen2011}. Viscoelastic effects are best investigated by introducing the Deborah number, defined as $De=\lambda f$, where $\lambda$ is the fluid relaxation time and $f$ is the nematode's beating frequency. The Deborah number represents the ratio of the time-scale of the fluid ``fading memory'' to the period of flow induced during undulatory swimming~\cite{Lauga2009}. Note that $\lambda=0$ for Newtonian fluids and $De\rightarrow \infty$ for an elastic solid. Figure~\ref{figure4}\emph{c} shows the nematode's swimming speed as a function of $De$. We find that the rapid increase in the nematode's swimming speed is not determined by $De$ (or fluid elasticity) but rather the transition from semi-dilute to concentrated fluid structure. 
\section{Discussion}
Our data show a rapid increase in the swimming speed of \emph{C. elegans} as the XG polymer concentration increases beyond 3000 ppm, which roughly corresponds to the transition from the solution semi-dilute to the concentrated regime. The main bulk rheological parameters such as viscosity factor $\mu_0$, power law index $n$, and fluid relaxation time (elasticity) $\lambda$ do not fully capture the observed phenomenon. This suggests that the polymer networks present in the concentrated polymeric fluids may be responsible for this enhancement in propulsion speed. 

In order to identify the mechanism behind the enhancement of swimming speed, we examine the normalized flow field streamlines shown in Fig.~\ref{figure1}\emph{c}~and~\emph{d} for nematodes swimming in semi-dilute (2000 ppm) and concentrated (4000 ppm) XG solutions (Non-normalized streamline videos available in SM). Both fields possess the similar flow structures. However, in the concentrated regime, we find a region of high relative velocity that is predominantly oriented normal to the nematode's body, whereas in the semi-dilute regime, this region is absent.  Note that simply modulating fluid viscosity would uniformly change velocities in the field in all directions; that is, the velocity fields would scale linearly with viscosity. This is clearly not that case in the concentrated regime, as the velocities normal to the nematode's mid section are larger relative to the tangential velocities. This indicates a more complex relationship between the swimmer and the networks in the polymer solution.  As discussed earlier (see Fig.~\ref{figure2}\emph{d}~and~\emph{e}), the fluid internal structure is sensitive to polymer concentration, particularly as the XG concentration increases from the semi-dilute to the concentrated regime~\cite{Rodd2000, Doi1988}. Our results indicates that the flow behavior of semi-dilute XG solutions is dominated by the hydrodynamic interaction among polymer molecules, while the flow behavior of concentrated solutions is dominated by molecule shape and dynamic properties of the polymer networks~\cite{Doi1988}. 

\begin{figure}[t!] 
    \begin{center}
      \includegraphics[width=0.47\textwidth]{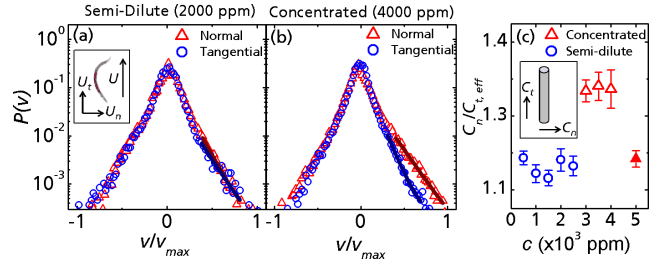}
       \caption{(Color Online) Probability distribution functions (PDF) of velocity components of tracer particles in {\bf{(a)}} semi-dilute and {\bf{(b)}} concentrated solutions. PDFs are computed at the same phase and are normalized by the total number of particles in the flow field. The blue circles ($\circ$) and red triangles ($\bigtriangleup$) represent the tangential and normal components of the velocity vectors with fitted exponential decay slopes. Inset: a schematic of nematode's velocity components. {\bf{(c)}} The effective drag coefficient ratios $C_n/C_{t,~eff}$ for semi-dilute (blue circles) and concentrated (red triangles) solutions. Inset: an illustration of drag coefficients for a slender cylinder.\label{figure5}}
    \end{center}
\end{figure}

We quantify the observed differences in the flow field by computing the probability distribution function (PDF) of the tracer particle velocities embedded in the flow. PDF plots are shown in Fig.~ \ref{figure5}\emph{a}~and~\emph{b} for a semi-dilute (2000 ppm) and a concentrated (4000 ppm) XG solution, respectively. The PDFs are decomposed into tangential and normal directions, which are computed with respect to the nematode's swimming direction. Here, the tangential direction corresponds to the nematode's swimming direction while the normal direction is perpendicular to the swimming direction. The area around the PDF's central peak, where velocity magnitudes $|v|$ are small, isotropic, and dominated by noise, represents the flow field far away from the swimming \emph{C. elegans}. The measurements of interest are at the PDF's tails, which correspond to regions of the flow field close to the nematode. 

Near the nematode's body, the velocity distributions exhibit exponential decay tails, indicating the dominant role of convection arising from the swimming motion. The solid lines represent an exponential fit of the decay tails of the form $P(v)=P(v)_{0}+Ae^{B \frac{v}{v_{max}}}$, where $P(v)$ is the probability density function (PDF) of velocity, $v$ and $v_{max}$ are the tracer particles' speed and maximum speed in the flow field, $A$ is a fitting coefficient, and $B$ is the decay slope. Results show that in the semi-dilute solution, both the normal and tangential velocity components show roughly the same decay slope: $B=-7.76,~-7.07$ respectively. A different trend, however, is found for the velocity distribution in concentrated solutions. While both tangential and normal components show exponential tails near the nematode's body (high velocities), we note an anisotropic distribution the decay rates of velocity components. The decay slopes for the tangential and normal directions are $B=-9.82,~-5.85$ respectively, indicating velocities in the tangential direction  decay faster at high velocities than those in the normal direction. This implies that for nematodes swimming in concentrated XG solution, there are regions where the fluid has a relative enhancement in momentum in the normal direction. At low Reynolds number, where fluid momentum is instantaneously dissipated, higher fluid speed and momentum along the normal direction require stronger driving force from the swimmer. Consequently, the nematode experiences an enhanced drag in the normal direction. Our data suggest that the nematode's motion is causing an anisotropic mechanical response in the fluid (in concentrated solutions), which enhances the normal viscous drag on the nematode and leads to the nematode's swimming speed enhancement. 

Next, we examine the physical properties of the XG solutions in the semi-dilute and concentrated regimes. As mentioned before, XG is a semi-rigid, rod-like polymer, which has a molecular weight (MW) of approximately $2 \times 10^{6}$, a hydrodynamic length of approximately 1.5~$\mu$m and a contour length of approximately 2.0~$\mu$m~\cite{Zirnsak1999}. At equilibrium, it has been established that for semi-rigid rod-like polymers, the fluid structure can transition from an isotropic fluid to an anisotropic nematic liquid crystal as the polymer concentration increases in the solvent~\cite{Tracy1992,Lim1984}. This transition occurs in the concentration regime~$\frac{4 (MW)}{N_A d \ell^{2}} \lesssim c \lesssim \frac{6 (MW)}{N_A d\ell^{2}},$ where $d$ and $\ell$ are the diameter and hydrodynamic length of the xanthan gum macromolecule respectively, $MW$ is its molecular weight, $N_A$ is Avogadro's number, and $c$ is the polymer concentration~\cite{Tracy1992}. Here, $d \approx 2$~nm, $\ell \approx 1.5$~$\mu$m, and $MW \approx 2 \times 10^6$~Da. For the xanthan gum solutions investigated here, this corresponds to a transition between concentrations of approximately 2900 and 4400 ppm. It is between these concentrations that we observe changes in the viscosity coefficient and power law index trends (Fig.~\ref{figure2}\emph{d}~and~\emph{e}) and in the nematode's swimming speed (Fig.~\ref{figure3}\emph{c}). It is believed that this transition occurs because these slender macromolecules can no longer freely rotate; rather, their interactions are becoming increasingly confined and crowded by adjacent molecules.  As the molecules become more crowded, the parallel component of the molecule's diffusion coefficient $D_\parallel$ becomes much greater than its perpendicular component $D_\perp$ because these rod-like molecules are more prone to slide past each other than to rotate~\cite{Dobrynin1995}.

The nematode's motion, however, disrupts the aforementioned equilibrium picture by introducing shear stresses to the system. The introduction of shear stresses in a rod-like polymeric solutions (e.g. XG fluids) causes molecules to align with the direction of shear due to their anisotropic drag~\cite{Song2006, Zirnsak1999}; that is, a rod-like molecule moves more easily along its length than normal to it. This alignment gives rise to the shear thinning viscosity behavior of XG solutions (Fig.~\ref{figure2}\emph{a}). The shear stresses introduced in the fluid by the nematode are also able to align these molecules, at least locally, and can lead to the formation of local nematic structures at high concentrations~\cite{Dobrynin1995}. The combination of these local nematic structures with the anisotropic diffusivity present only in the concentrated regime gives rise to the enhanced drag in the normal direction experienced by the nematode (Fig.~\ref{figure5}\emph{b}).

In order to elucidate the mechanism discussed above, i.e. the relative enhancement in drag in the normal direction, we estimate the drag coefficient ratio ${C_n}/{C_t}$ using resistive force theory~(RFT)~\cite{Gray1955} along with our kinematic data. We note that RFT is only valid for Newtonian fluids and its used here is solely to illustrate a possible mechanism. The ratio of the nematode's length ($\approx$~1 mm) to its diameter ($\approx$~80~$\mu$m) is approximately 12. Hence, in the limit of low $Re$, \emph{C. elegans} may be treated as a slender body moving in a viscous fluid. An expression for the swimming speed $U$ for an undulating filament  was proposed by Gray and Hancock~\cite{Gray1955} and is given as~$U=2\pi^{2}\left(\frac{f^2 A^2}{\nu_p}\right)\left(\frac{C_n}{C_t}-1\right)$ where $\nu_p$, $A$, and $f$ are the wave speed, beating amplitude, and frequency respectively. The quantity $C_n/C_t$ is the drag coefficient ratio where $C_n$ and $C_t$ are normal and tangential drag coefficients~\cite{Gray1955}. The presence of polymer networks in the fluid can affect this drag coefficient ratio~\cite{Francois2008}, and we use the above expression to approximate an \textit{effective} drag coefficient ratio, $C_n/C_{t,~eff}$. While these estimates cannot be quantitatively compared to those for a Newtonian fluid, they can provide valuable insight into the degree of anisotropy present in the fluid micro-structure.

Figure~\ref{figure5}\emph{c} shows the ratio $C_n/C_{t,~eff}$ as a function of XG concentration estimated by incorporating experimentally measured kinematics (i.e. $U$,  $\nu_p$, $A$, and $f$)~\cite{Gray1955}. We find a rapid increase in $C_n/C_{t,~eff}$ as the polymer concentration increases beyond 3000 ppm, where the solutions transition from the semi-dilute to the concentrated regime (Fig.~\ref{figure2}\emph{d}~and~\emph{e}). This suggests a relative increase of $C_n$ compared to $C_t$ for polymer concentration $> 3000$ ppm. This relative enhancement in the normal component of the drag coefficient $C_n$ is corroborated by measurements of the velocity fields~(Fig.~\ref{figure1}\emph{c}~and~\emph{d}) and the PDFs of their velocity distributions~(Fig.~\ref{figure5}\emph{a}~and~\emph{b}).

\section{Conclusion}
We have investigated the swimming behavior of the nematode \emph{C. elegans} in semi-dilute and concentrated solutions of a rod-like polymer (XG). We find a rapid increase in the nematode's swimming speed as the polymer concentration increases (Fig~\ref{figure3}\emph{c}). This sudden increase in swimming speed occurs near the solution's transition from the semi-dilute to concentrated regime. We show that this increase in swimming speed is most likely related to the anisotropic response of the fluid microstructure to applied stress due to the nematode's swimming motion. In short, the undulatory swimming motion of \emph{C. elegans} induces a structural anisotropy which leads to an increase in the \textit{effective} drag coefficient ratio $C_n/C_t$ and an enhancement in swimming speed $U$ (Fig.~\ref{figure5}\emph{c}). 

Experimentally measured velocity fields corroborate with the proposed mechanism. The PDFs of the velocities of tracer particles show important differences between semi-dilute and concentrated solutions (see Fig.~\ref{figure5}\emph{a}~and~\emph{b}). While for the semi-dilute case both the tangential and normal velocity components collapse onto one another, we find a sharp difference between the two components at high velocities for nematodes swimming in concentrated solutions. In particular, we find that the tangential velocity distribution has a faster decay at high velocities than the normal direction, which indicates a relative enhancement in the normal component of momentum. 

Our finding serves as a step towards understanding the locomotion of organisms in highly structured fluid environments such as human tissues, gels, and mucus. Such understanding can be important, for example, in modifying mucous systems to fend off bacterial infection, to treat human fertility disorders by altering the cervical fluid environment, and to better maintain ecological systems. 
\acknowledgments{
We thank T. Lamitina for providing \textit{C. elegans} as well as N. Keim and J. Sznitman for fruitful discussions. This work was supported by NSF CAREER (CBET) 0954084.}
\bibliographystyle{eplbib}
\bibliography{AnisotropicMedia}
\end{document}